\documentclass{aa}
\usepackage{graphicx}
\usepackage{txfonts}
\def\sbs{SBS\,1150+599A}
\def\png{PN\,G\,135.9+55.9}

\begin{document}

         \title{The kinematics of the most oxygen-poor
         planetary nebula \png}

       \subtitle{}
       \titlerunning{The kinematics of \png}

       \author{M. G. Richer
               \inst{1}
               \and
               J. A. L\'opez
               \inst{1}
               \and
               W. Steffen
               \inst{1}
               \and
               G. Tovmassian
               \inst{1}
               \and
               G. Stasi\'nska
               \inst{2}
               \and
               J. Echevarr\'\i a
               \inst{3}
       }

       \offprints{Michael Richer}

       \institute{Observatorio Astron\'omico Nacional, Instituto de
                  Astronom\'\i a, UNAM, P.O. Box 439027, San Diego,
                  CA, USA 92143-9027 \\
                  \email{\{richer, jal, wsteffen, gag\}@astrosen.unam.mx}
           \and
                  LUTH, Observatoire de Meudon, 5 Place Jules Janssen,
                  F-92195 Meudon Cedex, France \\
                  \email{grazyna.stasinska@obspm.fr}
           \and
                  Instituto de Astronom\'\i a, UNAM, Apartado
                  Postal 70-264, 04510 M\'exico, D.F., M\'exico \\
                  \email{jer@astroscu.unam.mx}
           }

       \date{Received ???; accepted ???}

       \abstract{
       \png\ is a compact, high excitation nebula
       that has been identified recently as the most oxygen-poor
       halo planetary nebula. Given its very peculiar
       characteristics and potential implications in the realms
       of stellar and Galactic evolution, additional data are needed to firmly
       establish its true nature and evolutionary history. Here we
       present the first long-slit, high spectral resolution
       observations of this object in the lines of H$\alpha$ and
       \ion{He}{ii}$\lambda$4686. The position-velocity data
       are shown to be compatible with the interpretation
       of \png\ being a halo planetary nebula.
       In both emission lines, we find
       the same two velocity components that characterize the
       kinematics as that of an expanding elliptical envelope.
       The kinematics is consistent with a prolate
       ellipsoidal model with axis ratio about 2:1, a radially decreasing
       emissivity distribution, a velocity distribution that is radial,
       and an expansion velocity of 30\,km/s for the bulk of the material.
       To fit the observed line profiles, this model requires an asymmetric
       matter distribution,
       with the blue-shifted emission considerably stronger than the
       red-shifted emission.  We find that the widths of the two velocity
       components are substantially wider than those expected due to
       thermal motions, but kinematic structure in the projected area
       covered by the slit appears
       to be sufficient to explain the line widths.
       The present data also rule out the possible
       presence of an accretion disk in the system that could have
       been responsible for a fraction of the H$\alpha$ flux, further
       supporting the planetary nebula nature of \png.
       }

      \maketitle

%

\section{Introduction}

\png\ is a recently-discovered planetary nebula in the Galactic
halo (Tovmassian et al. \cite{tovmassianetal2001}; initially known
as \sbs).  Its spectrum is quite unusual for a planetary nebula,
presenting only the Balmer lines of hydrogen,
\ion{He}{ii}$\lambda\lambda$4686,5411, and very weak
[\ion{O}{iii}]$\lambda$5007 ($\sim 3\%$ of H$\beta$) in the
4000-7000\AA\ interval. A photoionization model analysis showed
that such a spectrum implies a strongly density bounded and
extremely oxygen-poor nebula ionized by a very hot star.
Subsequent spectroscopy and imaging confirmed the weakness of the
collisionally-excited metal lines, the high excitation, and the
low oxygen abundance, around 1/500 of the solar value (Richer et
al. \cite{richeretal2002}). These data also showed that the object
is a point source in the continuum emission, but extended in
emission lines, and suffers little reddening.  Very puzzlingly,
these data revealed the possibility that the $\mathrm
H\alpha/\mathrm H\beta$ ratio is variable. Additional imaging and
spectroscopic observations by Jacoby et al.(\cite{jacobyetal2002})
generally confirmed these results, though they found
[\ion{Ne}{iii}]$\lambda$3869 and
[\ion{Ne}{v}]$\lambda\lambda$3346,3426 lines stronger by an order
of magnitude.

\png\ is such an important object because it has, by an order of
magnitude, the lowest oxygen abundance known among planetary nebulae
and will directly affect our understanding of several issues. Its study
will have important implications for the understanding of the synthesis
and mixing of chemical elements in metal-poor stars. \png\ may also
have important implications for the early evolution of the Milky Way
halo. Once this object is well-understood, it should also provide an
important limit upon the pre-galactic helium abundance. The
evolutionary history of \png, particularly once a reliable distance
becomes available, would provide a useful constraint upon the late
stages of evolution for metal-poor stars, e.g., the time scale during
pre-white dwarf evolution. More generally, kinematic studies of
planetary nebulae are important for understanding the production of
planetary nebulae from stellar populations (e.g., Stasi\'nska \&
Tylenda \cite{stasinskatylenda1994}; Richer et al.
\cite{richeretal1997}; Stasi\'nska et al. \cite{stasinskaetal1998};
Stanghellini \& Renzini \cite{stanghellinirenzini2000}; Gesicki \&
Zijlstra \cite{gesickizijlstra2000}).  A better understanding of the
production of planetary nebulae impacts a wide range of other
processes, including stellar death rates, white dwarf birth rates, the
mass of interstellar gas in galaxies without star formation, and the
chemical enrichment of galaxies in helium, carbon, nitrogen, and
s-process elements, among others.  While \png\ will not be determinant
for these latter studies, it will serve as a useful probe of a poorly
populated region of parameter space.

Given the limited morphological information available for \png\ (Richer
et al. \cite{richeretal2002}; Jacoby et al. \cite{jacobyetal2002}), a
study of its kinematics has the more immediate use of helping to
establish its nature and structure. We report the first high spectral
resolution observations that we use to study the kinematics of \png. In
Sect. 2 we describe the observations we have undertaken. In Sect. 3 we
present the results and their interpretation.  We find that \png\ is
best explained as an expanding elliptical envelope.  In Sect. 4 we
discuss how these results affect our understanding of \png.

\section{Observations and Reductions}

\begin{figure*}
\includegraphics[width=18cm]{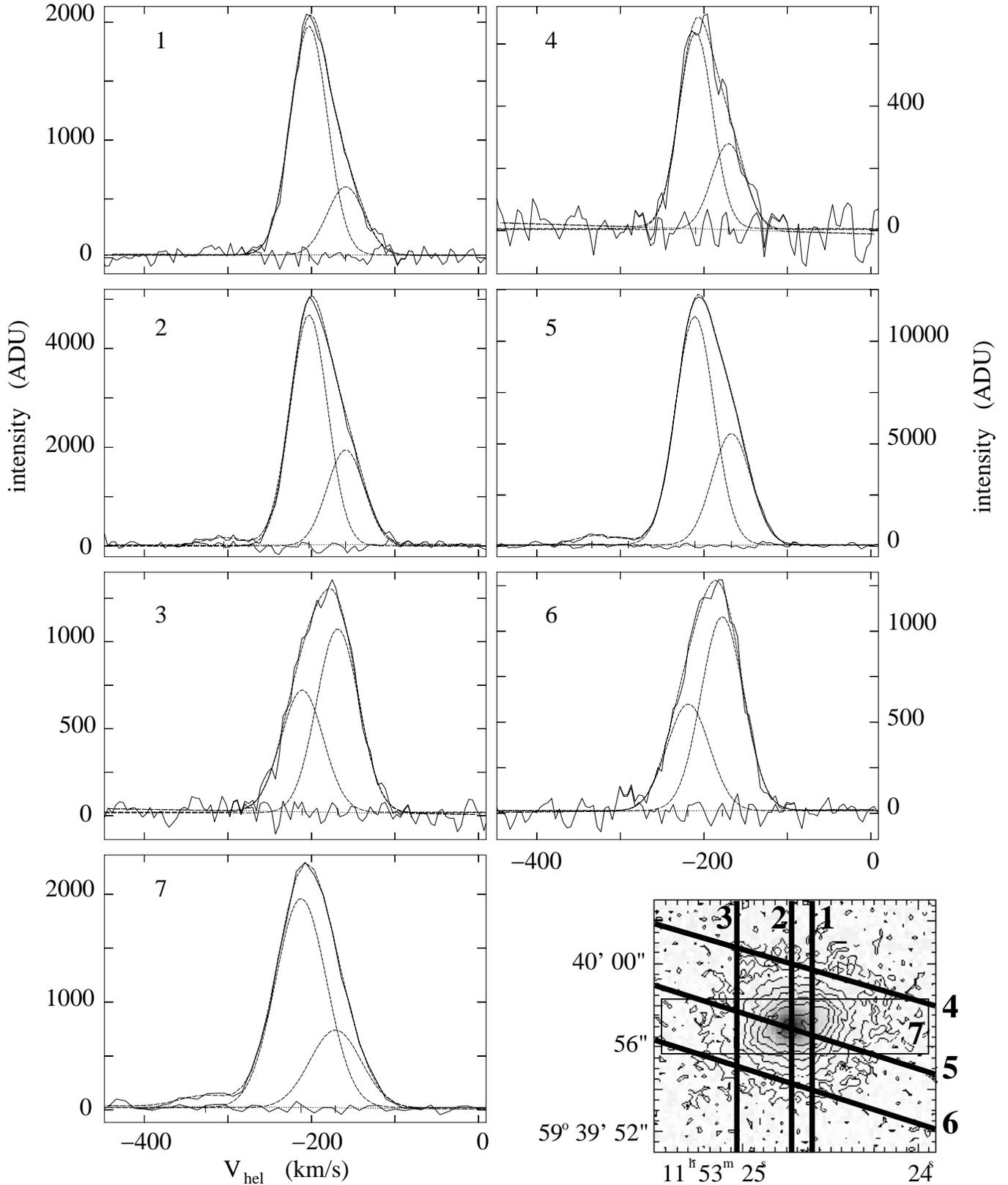}
\caption{This figure presents the H$\alpha$ line profiles for the
seven slit positions observed.  The numbers in the upper left
corners of each profile correspond to the slit positions shown in
the lower right image (H$\alpha$ contours, continuum grayscale; N is
up and E to the left).
The slit width for the MES observations
(slits 1-6) is the same as indicated for the
REOSC echelle observation (slit 7).  For each profile, the fit and
residuals are shown (from IRAF's splot) along with the
data.  The individual gaussian components at H$\alpha$,
created using IRAF's artdata.mk1dspec, are also superposed.
The wavelength scale (horizontal axis) is
plotted in units of the heliocentric radial velocity.  The faint
line visible to the blue of H$\alpha$ in slits 2, 5, and 7 is
\ion{He}{ii}$\lambda$6560.
}
\label{lineproffig}
\end{figure*}

High spectral resolution observations of \png\ were obtained at the
Observatorio Astron\'omico Nacional in San Pedro M\'artir, Baja
California, Mexico (SPM) with two different spectrometers at the f/7.9
Cassegrain focus of the 2.1m telescope.  The positions observed are
shown in the lower right panel of Fig. \ref{lineproffig}.

First, \png\ was observed on 8 and 10 January 2002 and 14 May 2002
using the Manchester echelle spectrometer (MES; Meaburn et al.
\cite{meaburnetal1984}).  The MES is a long slit, echelle spectrometer
that has no cross-dispersion. Instead, a narrow-band filter was used to
isolate the 87th order containing the H$\alpha$ nebular emission line.
A 150\,$\mu$m wide slit was used for the observations. Coupled with a
SITe $1024\times 1024$ CCD with 24\,$\mu$m pixels binned $2\times 2$,
the resulting spectral and spatial resolutions are 0.1\,\AA/pix
(equivalent to 11\,km/s for 2.5\,pix FWHM) and $1\farcs9$,
respectively. The spectra were calibrated to $\pm 1.0$\,km/s accuracy
using exposures of a ThAr lamp taken immediately after every object
exposure.  For the January 2002 observations, the slit was oriented
north-south and positioned at the positions 1-3 indicated in the lower
right panel of Fig. \ref{lineproffig}. Four 30 minute exposures were
obtained, one each at slit positions 1 and 3 and two at slit position 2
(Fig. \ref{lineproffig}).  For the May 2002 observations, six 30 minute
exposures were obtained with the slit oriented at a position angle of
75$^\circ$. Four of these exposures were at slit position 5 and single
exposures were obtained at slit positions 4 and 6 (Fig.
\ref{lineproffig}).  All of these MES spectra were bias-subtracted,
then extracted to one-dimensional spectra, calibrated in wavelength,
corrected to heliocentric velocities, and finally co-added for the
observations of slits 2 and 5.

Second, \png\ was observed on 19 January 2002 using the SPM REOSC
echelle spectrometer.  This is a conventional, cross-dispersed echelle
spectrometer.  Coupled with the same SITe CCD as the detector, the
resulting spectral resolution was 0.16\,\AA/pix at
\ion{He}{ii}$\lambda$4686 (equivalent to a velocity resolution of
26\,km/s), but substantially worse at H$\alpha$ due to camera de-focus
(see Table \ref{kintable}). The $2\arcsec \times 13\arcsec$ slit was
oriented east-west, as shown in Fig. \ref{lineproffig}. Two 30 minute
exposures of \png\ were obtained. The spectra were bias-subtracted,
co-added, extracted to one-dimensional spectra, calibrated in
wavelength, and finally corrected to heliocentric velocity.

For all of the spectra, the radial velocities and line widths were
measured using a locally-implemented software package (McCall et
al. \cite{mccalletal1985}).  This software fits the line profiles
with sampled gaussian profiles, fitting the systemic velocity, the
line intensities, and the line widths automatically.  In almost
all of the slit positions, a single gaussian was unable to
reproduce the line profiles satisfactorily due to the profile
asymmetry, so profiles composed of two gaussian components were
adopted.  To determine the optimal line separation, the line
profiles were fit with emission components whose separation was
varied iteratively until the standard deviation of the residuals
about the fitted profile reached a minimum.  In no case was there
any clear evidence for more than two velocity components in the
line profiles. In Fig. \ref{lineproffig}, the H$\alpha$ line
profiles are shown with the fitted profile and residuals
superposed.  The two velocity components were constrained to have
the same width.  For slits 3 and 6, it was possible to fit the
line profile acceptably with a single, broad gaussian component,
though these fits had larger residuals than the two component fits
given in Table \ref{kintable}.  Furthermore, the two component
fits for slits 3 and 6 are otherwise very similar to those found
at all other slit positions, which suggests that they are a
reasonable interpretation of the line profiles.

The decomposition of the line profiles as the sum of two gaussians
was initially motivated by simplicity.  However, we note that we
can fit line profiles in different positions within the nebula
where the profile shapes are different (Fig. \ref{lineproffig})
using the sum of two gaussians and that the properties of these
two components vary continuously and coherently as a function of
position. For these reasons, we interpret these two components as
representing the expansion of the two sides of the object.
However, spectroscopy with higher resolution, both spatially and
spectrally, will be required to determine whether there are indeed
two distinct gaussian components.

\begin{table}
\caption[]{kinematic data for \png} \label{kintable}
\[
\begin{tabular}{lcccc}
\hline \noalign{\smallskip} slit\,$^{\mathrm a}$ &
$V_\mathrm{blue}\,^{\mathrm b}$ & $\Delta
V_\mathrm{red}\,^{\mathrm c}$ & FWHM\,$^{\mathrm d}$ &
$I(\mathrm{blue})/I(\mathrm{red})$ \\
     & (km/s) & (km/s) & (km/s) & \\
\noalign{\smallskip}\hline\noalign{\smallskip} 1 & $-9.6\pm 0.5$
& $43.9\pm 1.4$ & $51.9\pm 1.0$ & $3.34\pm 0.20$ \\ 2 & $-9.7\pm
0.3$  & $43.9\pm 1.4$ & $52.3\pm 0.5$ & $2.41\pm 0.05$ \\ 3 &
$-17.7\pm 1.4$ & $42.5\pm 1.4$ & $59.4\pm 2.0$ & $0.66\pm 0.06$ \\
4 & $-16.6\pm 1.5$ & $39.3\pm 1.4$ & $47.6\pm 2.9$ & $2.28\pm
0.31$ \\ 5 & $-17.5\pm 0.2$ & $43.8\pm 1.4$ & $54.4\pm 0.3$ &
$2.05\pm 0.03$ \\ 6 & $-25.7\pm 1.6$ & $41.1\pm 1.4$ & $59.7\pm
2.4$ & $0.55\pm 0.06$ \\ 7(H$\alpha$) & $-19.2\pm 1.4$ & $41.1\pm
2.0$ & $72.1\pm 1.6$ & $2.69\pm 0.35$ \\ 7(4686) & $-24.1\pm 3.0$
& $43.7\pm 4.5$ & $53.6\pm 5.3$ & $1.90\pm 0.38$ \\
\noalign{\smallskip}\hline
\end{tabular}
\]
\begin{list}{}{}
\item[$^{\mathrm a}$] All values are for the H$\alpha$ line,
except for slit 7 where results for both H$\alpha$ and
\ion{He}{ii}$\lambda$4686 are given (see Fig. \ref{lineproffig}).
\item[$^{\mathrm b}$] This is the radial velocity of the
blue-shifted component with respect to the heliocentric systemic
radial velocity of $-193.3\pm 1.3$\,km/s, the latter based upon the
intensity-weighted average velocities for slits 2 and 5.
\item[$^{\mathrm c}$] This is the velocity difference between the
blue- and red-shifted velocity components.
\item[$^{\mathrm d}$] This is the width (FWHM) of the two velocity
components.  Both components were constrained to have the same
width.
\end{list}
\end{table}

\section{Results}

In Table \ref{kintable}, we list the results of our line profile
analyses.  These results include the radial velocity of the blue
velocity component, the separation of the blue and red velocity
components, the width of two components, and the ratio of the
intensities of the two components.  The radial velocity of the blue
component is given relative to the heliocentric systemic velocity of
\png. We adopted the average of the intensity-weighted mean velocities
for slits 2 and 5 as the systemic velocity.  The intensity-weighted
mean velocities for slits 2 and 5 were computed as the average velocity
of the blue- and red-shifted components, weighted according to their
respective intensities. The resulting heliocentric systemic velocity
for \png\ was found to be $-193.3\pm 1.3$\,km/s.  The width of the
velocity components tabulated in column 4 of Table \ref{kintable} is
the full width at half maximum (FWHM) of the individual velocity
components (both components having been constrained to have the same
width). The tabulated uncertainties are those derived from the fit to
the line profiles, except in the case of the uncertainty in the line
separation, which is taken to be 1.5 times the step size used to search
for the minimum in the standard deviation of the fit.

As is evident from the line profiles (Fig. \ref{lineproffig}), the
sequences of slit positions 1-3 and 4-6 are sequences in which the
intensity of the blue velocity component decreases relative to the
red component.  The relative intensity of the red component
increases towards the east and south.  Given that these slit
positions include a significant fraction of the object, the
variation of the relative intensities of the blue- and red-shifted
components could be more pronounced than Table \ref{kintable}
implies.

Likewise, the sequences of slit positions 1-3 and 4-6 are sequences in
which the velocity of the blue component, $V_\mathrm{blue}$, also
increases in the last position.  The largest velocities of approach of
the blue component are found for slits 3 and 6, in the east and
south-east, respectively.

From Table \ref{kintable}, it is clear that the separation of the two
velocity components is approximately constant at 39-44\,km/s. It is
also noteworthy that the separation of the velocity components is the
same in H$\alpha$ and \ion{He}{ii}$\lambda$4686. The entire nebular
envelope is therefore characterized by an average apparent expansion
velocity of 20-22\,km/s.

Finally, we note that the widths of the two velocity components at all
of the slit positions are substantially wider than those expected for a
plasma at the expected temperature of 30,000\,K (Richer et al.
\cite{richeretal2002}).  At this temperature, the line widths are
expected to be 37\,km/s and 19\,km/s at H$\alpha$ and
\ion{He}{ii}$\lambda$4686, respectively (equation 2-243 from Lang
\cite{lang1986}). Instead, line widths of approximately 55\,km/s are
found. (We ignore the H$\alpha$ line width in slit 7 due to the
spectrometer de-focussing.)  The large observed width of the lines
could arise from several causes. Given that the slit width covers a
large fraction of the object, it is likely that it includes
considerable kinematic structure, which would broaden the lines.

\section{Discussion}

Here, we develop a spatiokinematic model that explains the kinematics
observed in \png.  Given the relatively low spatial resolution of both
our high-resolution spectroscopy and extant imaging (Richer et al.
\cite{richeretal2002}; Jacoby et al. \cite{jacobyetal2002}), our model
will certainly not be definitive, but it should be sufficient to
establish the large-scale features.  A definitive model will have to
await the availability of high-resolution imaging.

We have used the modeling code SHAPE (Steffen et al.
\cite{steffenetal1996}) to investigate the basic structure of \png.
SHAPE produces long slit spectra from a simple mathematical model of
the emissivity and velocity field of three dimensional objects. These
can be constructed as an arbitrary combination of individual model
structures.  The objects are placed at any orientation in a cubic
volume, which is divided into elementary cells of uniform emissivity.
The image and spectrum is calculated for each object and combined with
the results of the others, taking obscuration into account, if
necessary.  The emission is integrated along different lines of sight
in order to produce an image. The long slit spectra are produced by
selecting those cells that are situated along the lines of sight
covered by the slit. The emission is then mapped in position along the
slit and in velocity.  The result is then convolved in space and
velocity according to the corresponding angular and spectral
resolution. Thermal broadening is approximated by decreasing the
velocity resolution appropriately.  One-dimensional line profiles may
then be obtained by integrating the position-velocity maps along the
spatial direction.  As an example application of SHAPE, see L\'opez et
al. (\cite{lopezetal1997}).

\begin{figure*}
\includegraphics[width=18cm]{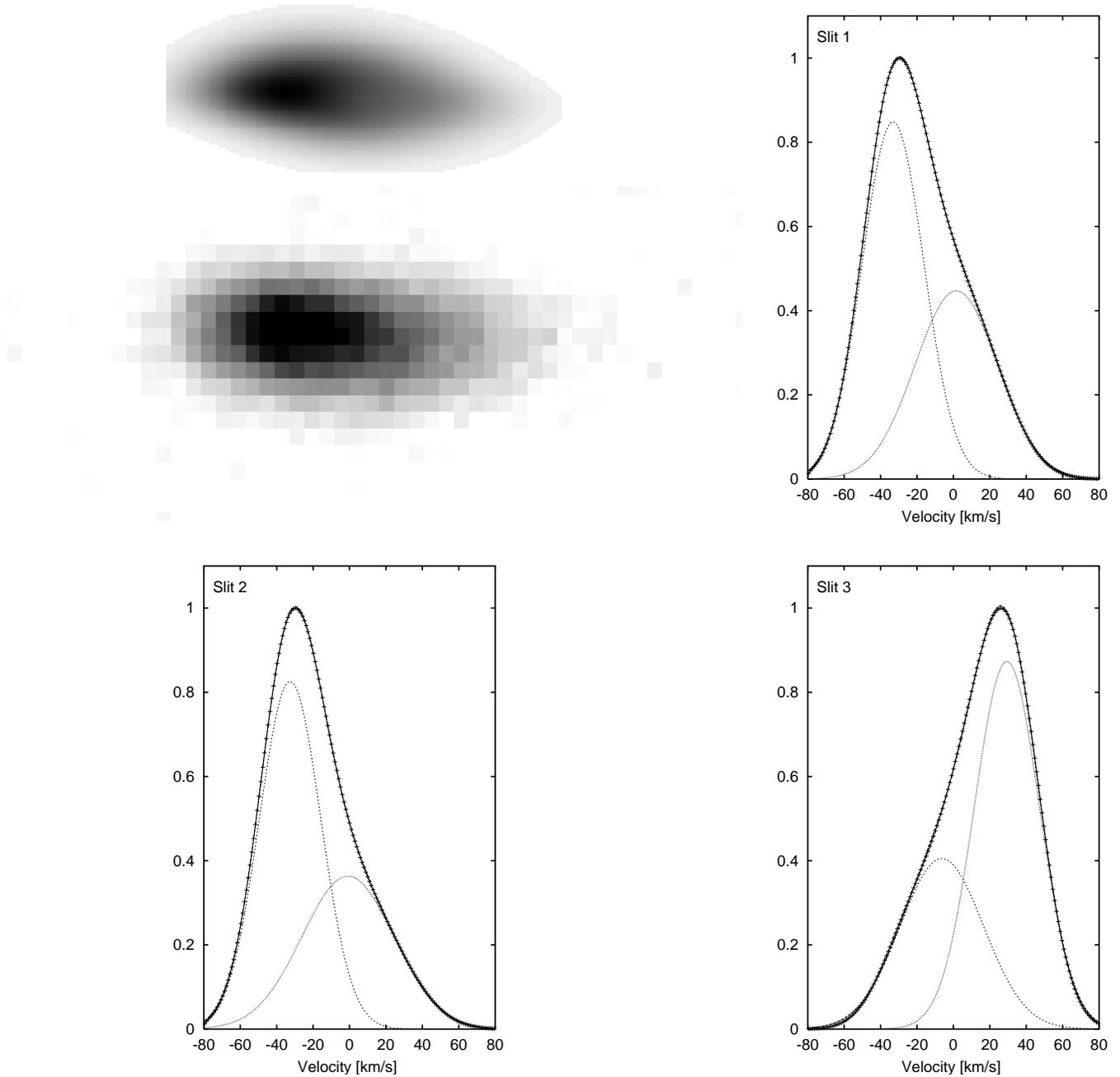}
\caption{Here, we present the results of the prolate elliptical model
(Model 2; see text).  The upper left panel presents the model
position-velocity diagram for slit 2 while the middle left panel
presents the observed position-velocity diagram.  At bottom left is the
integrated line profile for slit 2 predicted by Model 2.  In the right
hand panels, the integrated line profiles for slit 1 and slit 3 are
presented in the top and bottom panels, respectively.  These line
profiles may be compared with those observed in Fig. \ref{lineproffig}.
The individual profiles of the blue- and red-shifted material are also
shown.  The velocities are all relative to the systemic velocity.}
\label{modproffig}
\end{figure*}

Given the modest spatial resolution of the current observations, very
detailed structural models are not warranted. However, available
imaging, particularly that of Jacoby et al. (\cite{jacobyetal2002}),
indicates that the structure is not spherically symmetric, but rather
elliptical. We calculated spectra for three different structures of
increasing complexity.  First, we constructed a spherically-symmetric
model with isotropic (radial) gaussian emissivity and velocity field
that is radial and uniform (Model 1). Second, as a generalization of
the isotropic model, we constructed a prolate elliptical model (\lq\lq
cigar"-shaped), with a gaussian emissivity distribution, where the FWHM
towards the poles is roughly twice as long as that in the equatorial
plane (Model 2). The velocity in Model 2 is still radial, but its value
in each direction is proportional to the FWHM of the emissivity
distribution. Additionally, the brightness of one side of this model
can be reduced arbitrarily compared to the other. Finally, we have
considered a slow or static, isotropic, faint halo combined with Model
2 (Model 3), which seems to be distinguished from the elongated
structure in the image by Jacoby et al. \cite{jacobyetal2002}.

We have evaluated slit positions 1, 2 and 3 (see Fig.
\ref{lineproffig}). The best results were found for the prolate
elliptical Model 2 with the following parameters: FWHM of volume
emissivity in (x, y, z) is (3.7, 3.7, 7.5) arc seconds, velocity field
is radial with ($V_x$, $V_y$, $V_z$) of (30, 30, 70) km/s on the axes,
with the position angle of z-axis at 290 degrees in the plane of the
sky and inclined with respect to the line of sight by 65 degrees.  In
the model presented in Fig. \ref{modproffig}, the red-shifted emission
has only 20\% of the intensity of the blue-shifted emission.  These
parameters should be considered indicative only, as a more extensive
search of parameter space would likely uncover additional models that
fit the data within the uncertainties. An isotropic emissivity
distribution (Model 1) does not fit the data. The addition of an
isotropic, faint halo (Model 3) does not significantly improve upon the
agreement found between observations and Model 2. Particularly in the
case of slit 2, the agreement between observations and Model 2 is
excellent.

We therefore adopt 30\,km/s as the expansion velocity for the bulk of
the matter in \png.  Note that this is significantly different from the
apparent expansion velocity determined by fitting two gaussians to the
line profile (Sect. 2), since the fit from Model 2 takes into account
projection effects.

Our modelling experiments clearly indicate that the matter distribution
is asymmetric.  In Fig. \ref{modproffig}, we present the integrated
line profiles for slit positions 1-3 from Model 2.  In this figure, we
also compare the position-velocity diagrams observed and predicted by
Model 2 for slit position 2. For Model 2, we found consistency with the
data only using asymmetric structures, with the red-shifted emission
considerably weaker than the blue-shifted emission, reflecting the
intensity ratios given in Table \ref{kintable}.  The most plausible
physical mechanisms for this emissivity asymmetry are internal
reddening and an asymmetric matter distribution.  Since the reddening
of both the nebula and central star are low (Tovmassian et al.
\cite{tovmassianetal2003}), internal reddening does not appear to be a
viable explanation for the asymmetry of the emission. It is therefore
likely that the matter distribution is asymmetric and that the
red-shifted emission is intrinsically weaker than the blue-shifted
emission.  It is also likely that high-resolution imaging will find a
more complex matter distribution than the simple one adopted in Model
2, something that could be easily incorporated in future work. Indeed,
even at the resolution of the H$\alpha$ image presented by Jacoby et
al. (\cite{jacobyetal2002}), \png\ shows substructure.

It is clear from the preceding modelling experiments that kinematic
structure within the projected area of the slit significantly affects
the interpretation of the line profiles. This kinematic structure
appears to account for the breadth of the two components we measure
from the profiles. The large range of velocities included by the slit
broadens the line profiles well beyond the thermal widths.

There are relatively few halo planetary nebulae known, and the
kinematics of only a fraction of these have been studied.  With the
exception of NGC 4361, halo planetary nebulae appear to have expansion
velocities in the 10-30\,km/s range (H4-1 and BB1: Sabbadin et al.
\cite{sabbadinetal1986}; IC 4997: Miranda et al.
\cite{mirandaetal1996}; PN G009.8-07.5: Jacoby et al
\cite{jacobyetal1998}).  NGC 4361 has a more complex, bi-polar
kinematics (V\'azquez et al. \cite{vazquezetal1999}).  Compared to
other halo planetary nebulae, the kinematics we find for \png\ does not
appear to be particularly atypical.

Considering the similar kinematics and line widths in both
H$\alpha$ and \ion{He}{ii}$\lambda$4686, it is evident that both
H$^+$ and He$^{2+}$ should occupy very similar nebular volumes.
This result agrees with the null detections of
\ion{He}{i}$\lambda$5876 in low resolution spectra and confirms
that the nebular envelope is very optically thin (Richer et al.
\cite{richeretal2002}; Jacoby et al. \cite{jacobyetal2002}).  The
same mechanism should then be responsible for the similar line
widths of both H$\alpha$ and \ion{He}{ii}$\lambda$4686.

One of the goals of the deep spectrum obtained at slit position 5 was
to investigate the possible existence of broad H$\alpha$ emission, $>
\mathrm{few}\times 100$\,km/s.  No such broad component was found, in
agreement with the results of Tovmassian et al.
(\cite{tovmassianetal2001}). This result rules out the possibility that
\png\ contains an accretion disk that emits a significant fraction of
the H$\alpha$ flux, even for the case of a face-on disk (e.g., Horne \&
Marsh \cite{hornemarsh1986}). Consequently, we can discard the
possibility that \png\ contains a close interacting binary with an
accretion disk that could introduce a distinct $\mathrm H\beta/\mathrm
H\alpha$ ratio or any sort of variability.

To summarize, the kinematics of \png\ allow us to impose important
constraints upon the structure of this most oxygen-poor planetary
nebula.  Globally, the kinematics can be interpreted as an expanding
elliptical envelope.  Our modelling experiments indicate that the
matter distribution is asymmetric, a prolate ellipsoid fitting best,
with the blue-shifted emission several times stronger than the
red-shifted emission.  The apparent line widths we find are
substantially larger than the thermal widths, though this appears to be
due to the kinematic structure projected within the area of the slit.
There is no indication of the presence of an accretion disk or a binary
companion. Finally, given the extreme nature of its spectrum, that the
kinematics of \png\ are similar to those of other halo planetary
nebulae is additional evidence in favour of it in fact being a
planetary nebula.

\begin{acknowledgements}

MGR, JAL, and GT thank G. Garc\'\i a, G. Melgoza, S. Monrroy, and
H. Riesgo for their able assistance with the observations at SPM.
We gratefully acknowledge financial support from CONACyT projects
32214-3 (JAL), 34521-E (GT), and 37214-E (MGR, GS) and DGAPA
projects IN100799 (MGR), IN114199 (JAL), and IN118999 (JE, GT).

\end{acknowledgements}

\end{document}